\documentclass[preprint2,emulateapj]{aastex}
\usepackage[]{natbib}

\newcommand{\msol}{M$_\odot$}

\slugcomment{Accepted for publication in ApJ}

\shorttitle{Near-infrared Linear Polarization of Ultracool Dwarfs}
\shortauthors{Zapatero Osorio et al.}

\begin{document}

\title{Near-infrared Linear Polarization of Ultracool Dwarfs}

\author{M$.$ R$.$ Zapatero Osorio}
\affil{Centro de Astrobiolog\'\i a (CSIC-INTA), Ctra$.$ Ajalvir km 4, E-28850 Torrej\'on de Ardoz, Madrid, Spain}
\email{mosorio@cab.inta-csic.es}

\author{V$.$ J$.$ S$.$ B\'ejar\altaffilmark{1}}
\affil{Instituto de Astrof\'\i sica de Canarias, C/$.$ V\'\i a L\'actea s/n, E-38205 La Laguna, Tenerife, Spain}
\email{vbejar@iac.es}

\author{B$.$ Goldman}
\affil{Max-Planck Institut f\"ur Astronomie, K\"onigstuhl 17, D-69117 Heidelberg, Germany  }
\email{goldman@mpia.de}

\author{J$.$ A$.$ Caballero}
\affil{Centro de Astrobiolog\'\i a (CSIC-INTA), P.O$.$ Box 78, E-28691 Villanueva de la Ca\~nada, Madrid, Spain}
\email{caballero@cab.inta-csic.es}


\author{R$.$ Rebolo\altaffilmark{1,2}, J$.$ A$.$ Acosta-Pulido\altaffilmark{1}, A$.$ Manchado\altaffilmark{1,2}, and K. Pe\~na Ram\'\i rez\altaffilmark{1}}
\affil{Instituto de Astrof\'\i sica de Canarias, C/$.$ V\'\i a L\'actea s/n, E-38205 La Laguna, Tenerife, Spain}
\email{rrl@iac.es, jaa@iac.es, amt@iac.es, karla@iac.es}

\altaffiltext{1}{Dept$.$ Astrof\'\i sica, Univ$.$ La Laguna, Tenerife, Spain}
\altaffiltext{2}{Consejo Superior de Investigaciones Cient\'\i ficas (CSIC), Madrid, Spain}


\begin{abstract}
We report on near-infrared $J$- and $H$-band linear polarimetric photometry of eight ultracool dwarfs (two late-M, five L0--L7.5, and one T2.5) with known evidence for photometric variability due to dust clouds, anomalous red infrared colors, or low-gravity atmospheres. The polarimetric data were acquired with the LIRIS instrument on the William Herschel Telescope. We also provide mid-infrared photometry in the interval 3.4--24 $\micron$ for some targets obtained with {\sl Spitzer} and {\sl WISE}, which has allowed us to confirm the peculiar red colors of five sources in the sample. We can impose modest upper limits of 0.9\,\%~and 1.8\,\%~on the linear polarization degree for seven targets with a confidence of 99\,\%. Only one source, 2MASS\,J02411151$-$0326587 (L0), appears to be strongly polarized ($P \sim 3\,\%$) in the $J$-band with a significance level of $P/\sigma_P \sim 10$. The likely origin of its linearly polarized light and rather red infrared colors may reside in a surrounding disk with an asymmetric distribution of grains. Given its proximity (66\,$\pm$\,8 pc), this object becomes an excellent target for the direct detection of the disk. 
\end{abstract}


\section{Introduction}

There is growing evidence that dwarfs of spectral types L and T show a large spread of near- and mid-infrared colors that cannot be explained by simple models (e.g., Marley et al$.$ \cite{marley10}). It is believed that gravity, metallicity, and distribution of dust clouds play a critical role in defining the atmospheric properties of L- and T-type sources (see Kirkpatrick \cite{kirk05} for a review). Various groups have monitored photometrically a number of L and T objects aimed at characterizing cloud patchiness in brown dwarfs (e.g., Gelino et al$.$ \cite{gelino02};  Koen \cite{koen04a}; Koen et al$.$ \cite{koen04b}; Artigau et al. \cite{artigau06}). Polarimetric observations at optical wavelengths have also been attempted to confirm the presence of dusty clouds in the objects' atmospheres (M\'enard et al$.$ \cite{menard02}: Zapatero Osorio et al$.$ \cite{osorio06}; Goldman et al$.$ \cite{goldman09}; Tata et al$.$ \cite{tata09}). The main results are that photometric variability (if detected) has a relatively small amplitude (typically below 0.1 mag in the $I$-band and infrared wavelengths), variability can be periodic (with modulations in agreement and disagreement with the expected rotation periods), non-periodic variability also appears quite commonly, and a few percentage of the L dwarfs show linear polarization degrees typically below 1\%~in the optical bands. All these observations support the presence of condensates in the form of dust clouds and the fast rotation of the dwarfs.

Among the known L dwarfs, there are some showing markedly redder near- and mid-infrared colors than expected for their spectral type . The origin of this property is not well understood. Burrows et al$.$ \cite{burrows06}, Leggett et al$.$ \cite{leggett07}, Burgasser et al$.$ \cite{burgasser08}, and Cruz et al$.$ \cite{cruz09} argued that low gravity causes red colors, but not all red L dwarfs necessarily have low surface gravities. An excess of metallicity and/or unusual condensate properties can also be the cause of a reddish nature. In Zapatero Osorio et al$.$ \cite{osorio10} we explored various scenarios for one particular very red L dwarf: G\,196$-$3\,B. We concluded that a low-gravity atmosphere with enshrouded upper atmospheric layers and/or a warm dusty disk/envelope provides the most likely explanations, the two of them consistent with an age below 300 Myr.

Imaging polarimetry provides a powerful tool for studying the properties of the atmospheric dust clouds and surrounding disks/envelopes. While the models by Sengupta \& Marley \cite{sengupta09,sengupta10} (and references therein) predict non-zero disk-integrated polarization with degrees quite below 1\%~for the L and T dwarfs, polarimetric observations of young stars with disks yield higher linear polarization intensities (e.g., Kusakabe et al$.$ \cite{kusakabe08}; Pereyra et al$.$ \cite{pereyra09}, and references therein). Here, we report on the first results of our polarimetric observations of ultracool dwarfs (spectral types later than M7) with peculiar colors and photometric variability or low gravity atmospheres.

\section{Target selection \label{targets}}

Our main targets are northern examples of ultracool dwarfs with spectral types late-M through T and hallmarks of infrared flux excesses or low-gravity atmospheres, and ultracool dwarfs with photometric variability reported in the literature. Additionally, all these objects must be bright in the $J$-band ($J \le 16.5$ mag) to obtain accurate polarimetric observations. Most targets were selected from the catalog of low-gravity L dwarfs by Cruz et al$.$ \cite{cruz09}, and from the numerous lists of young late-M and L-type brown dwarf member candidates of well known star-forming regions such as Taurus and Upper Scorpius. Below we provide a brief summary on the properties of the eight science targets listed in Table~\ref{log}, which represents a subsample of our longer target list. 

2MASS\,J01365662$+$0933473 (J0136$+$09) is the second brigh\-test T dwarf discovered so far (Artigau et al$.$ \cite{artigau06}). It is classified as a T2.5 dwarf located at a photometric distance of $\sim$6.4\,pc. Artigau et al$.$ \cite{artigau09} reported the near-infrared photometric variability of this object. The resulting light curve showed a periodic modulation with a period of $\sim$2.4\,h, a peak-to-peak amplitude of 50\,mmag, and significant variations from night to night. The authors discussed that the measured periodicity is related to the dwarf rapid rotation and that the light curve variations are attributable to evolution of atmospheric structures and/or differential rotation. Artigau et al$.$ \cite{artigau09} suggested that the atmosphere of J0136$+$09 contains grain-bearing cloudy regions. The age of this T dwarf is unconstrained.

2MASS\,J22443167$+$2043433 (J2244$+$20) is an L6.5 (optical classification, Kirkpatrick et al$.$ \cite{kirk08}) or an L7.5 (near-infrared typing, Knapp et al$.$ \cite{knapp04}) field dwarf displaying weak K\,{\sc i} lines in the near-infrared wavelengths  (McLean et al$.$ \cite{mclean03}). This effect is thought to be associated to surface gravity and/or metallicity. On the contrary, its optical spectrum does not show significant differences compared to other L dwarfs whose spectral types lie within the broad range L5.5--L9. Knapp et al$.$ \cite{knapp04}, Golimowski et al$.$ \cite{golimowski04} and Leggett et al$.$ \cite{leggett07} reported that the $J-H$, $H-K_s$, $K_s-L'$, and $L'-M'$ colors of J2244$+$20 are significantly red, which the authors attributed to condensate clouds more optically thick than usual. A low-amplitude, periodic photometric variability was observed at 4.5\,$\mu$m by Morales-Calder\'on et al$.$ \cite{morales06}, suggesting that the object's rotation period is about 4.6 h. In Zapatero Osorio et al$.$ \cite{osorio05} we detected some linear polarization in the $I$-band for this source, which was not confirmed by the recent, more accurate measurements of Goldman et al$.$ \cite{goldman09}. 

2MASS\,J02411151$-$0326587 (J0241$-$03), 2MA\-SS\,J03552337$+$1133437 (J0355$+$11) and 2MASS J10224821$+$5825453 (J1022$+$58) are low-gravity dwarfs with spectral types L0$\gamma$, L5$\gamma$ and L1$\beta$ according to Cruz et al$.$ \cite{cruz09}. Their optical spectra display features typical of objects younger than 300\,Myr such as notably strong VO and weak FeH molecular absorption and weak Na\,{\sc i} and K\,{\sc i} doublets. J0241$-$03 has no lithium feature in its spectrum (pseudo-equivalent width pEW $\le$ 2\,\AA) indicating either a very low gravity and extremely young age at which the line cannot be seen (Kirkpatrick et al$.$ \cite{kirk08}) or an efficient depletion of this element by nuclear reactions (see Rebolo et al$.$ \cite{rebolo92}), an age of the order of a few hundred Myr,  and a mass likely larger than 0.06\,\msol~(Chabrier et al$.$ \cite{chabrier00}). As illustrated in Fig.~6 of Cruz et al$.$ \cite{cruz09}, this object displays a $J-K_s$ color markedly redder than expected for its spectral classification; its {\sl Spitzer} mid-infrared indices (3--8\,$\mu$m) provided by Luhman et al$.$ \cite{luhman09} also appear redder than those of other L0 dwarfs in the field (see Fig.~\ref{color}). The origin of such photometric property remains unexplained, although Zapatero Osorio et al$.$ \cite{osorio10} speculated on different scenarios (see below). The L5-type dwarf J0355$+$11 shares a similar red condition in its $J-K_s$ color as J0241$-$03. J0355$+$11 has preserved lithium in its atmosphere suggesting a mass likely around or below 0.06\,\msol~(Chabrier et al$.$ \cite{chabrier00}), and confirming its substellarity. Blake et al$.$ \cite{blake10} found that J0355$+$11 has a small projected rotational velocity ($v$\,sin\,$i$ $\sim$ 12\,km\,s$^{-1}$), which contrasts with the typically high rotation ($\ge$\,20\,km\,s$^{-1}$) of many other field L dwarfs (Zapatero Osorio et al$.$ \cite{osorio06}; Reiners \& Basri \cite{reiners08}). Bernat et al$.$ \cite{bernat10} announced the detection of a candidate companion at a separation of 82.5 mas from this source. The L1 dwarf J1022$+$58 has no lithium (pEW $\le$ 1\,\AA, Cruz et al$.$ \cite{cruz09}), a low to moderate projected rotational velocity ($v$\,sin\,$i$ = 11.8--15 km\,s$^{-1}$), and significant, variable H$\alpha$ emission (Schmidt et al$.$ \cite{schmidt07}; Reiners \& Basri \cite{reiners08}; Blake et al$.$ \cite{blake10}). In contrast to other L dwarfs in our sample, J1022$+$58 does not show a red $J-K_s$ color. Interestingly, Blake et al$.$ \cite{blake10} discussed that due to its large spatial velocity, J1022$+$58 may be part of a distinct, old (and possibly low metallicity) thick disk population. 

G\,196$-$3\,B is a lithium-bearing, substellar proper motion companion at 16\arcsec~from the young M2.5 star G\,196$-$3\,A (Rebolo et al$.$ \cite{rebolo98}). It was classified as a low-gravity L3$\beta$ dwarf (Cruz et al$.$ \cite{cruz09}), and its photometric and spectroscopic properties from the visible to the mid-infrared wavelengths are widely reported in the literature (Mart\'\i n et al$.$ \cite{martin99}; Basri et al$.$ \cite{basri00}; Kirkpatrick et al$.$ \cite{kirk01,kirk08}; Gizis et al$.$ \cite{gizis02}; Mohanty \& Basri \cite{mohanty03}); McGovern et al$.$ \cite{mcgovern04}; McLean et al$.$ \cite{mclean07}; Allers et al$.$ \cite{allers07}; Bihain et al$.$ \cite{bihain10}). G\,196$-$3\,B exhibits markedly red colors at all wavelengths from 1.6 to 24\,$\mu$m. In Zapatero Osorio et al$.$ \cite{osorio10} we discussed various physical scenarios to account for the reddish nature of G\,196$-$3\,B, and concluded that a low-gravity atmosphere with enshrouded upper atmospheric layers and/or a warm dusty disk/envelope provides the most likely explanations. From kinematic considerations, the system G\,196$-$3 is a likely member of the Local Association and its age is constrained to the interval 20--100 Myr. At this young age, G\,196$-$3\,B would have a mass of 0.012--0.025\,\msol, near the planet--brown dwarf boundary. Gizis et al$.$ \cite{gizis02} reported a relatively low projected rotational velocity ($v$\,sin\,$i$ $\sim$ 15\,km\,s$^{-1}$) for this substellar source.

UScoCTIO\,128 (USco\,128) and UScoCTIO\,132 (USco\,132) are two M7-type member candidates of the 10-Myr Upper Scorpius star-forming association identified by Ardila et al$.$ \cite{ardila00}. USco\,128 has lithium detection confirming its young age and substellar nature, an upper limit on its projected rotation velocity ($v$\,sin\,$i$ $\le$ 5\,km\,s$^{-1}$), strong (and possibly variable) H$\alpha$ emission and emission of He\,{\sc i} and Ca\,{\sc ii} suggesting accretion events (Jayawardhana et al$.$ \cite{jayawardhana02}; Muzerolle et al$.$ \cite{muzerolle03}; Mohanty et al$.$ \cite{mohanty05}; Herczeg et al$.$ \cite{herczeg09}). An infrared flux excess compatible with an accretion disk was detected in USco\,128 by Jayawardhana et al$.$ \cite{jayawardhana03} and Scholz et al$.$ \cite{scholz07}. Using radial velocity measurements Kurosawa et al$.$ \cite{kurosawa06} identified this M7 dwarf as a binary candidate, which constrasts with the direct high-resolution imaging observations of Kraus et al$.$ \cite{kraus05}. Despite of the $L$-band flux excess observed by Jayawardhana et al$.$ \cite{jayawardhana03}, USco\,132 lacks lithium and has a dwarflike K\,{\sc i} absorption in its optical spectrum, indicating that it is most likely not member of the young association (Muzerolle et al$.$ \cite{muzerolle03}).

\section{Observations and results}
\subsection{Mid-infrared {\sl Spitzer} photometry}
Targets J0241$-$03, G\,196$-$3\,B, J2244$+$20  and USco\,128 have published mid-infrared photometry obtained with the various instruments onboard {\sl Spitzer}. We compiled these data and the corresponding references in Table~\ref{spitzer}. To complement the infrared photometry of the sample between 2.2 ($K$-band) and 24~$\mu$m, we downloaded public images at 3.55, 4.493, 5.731, and 7.872 $\mu$m acquired with the Infrared Array Camera (IRAC; Fazio et al$.$ \cite{fazio04}), and images at 23.675\,$\mu$m taken with the Multiband Imaging Photometer for {\sl Spitzer} (MIPS; Rieke et al$.$ \cite{rieke04}) when available from the {\sl Spitzer} Heritage Archive. The corresponding Astronomical Observation Request (AOR) identification numbers and observing dates are all listed in Table~\ref{spitzer}. The new data were acquired for J0136$+$09 (IRAC$+$MIPS), J0241$-$03 (MIPS), J1022$+$58 (IRAC), and J0355$+$11 (IRAC). Raw data were reduced with the {\sl Spitzer} Science Center S18.7.0 (IRAC) and S16.1.0 (MIPS) pipelines, which produced processed images with plate scales of 0\farcs6 pix$^{-1}$ (IRAC) and 2\farcs45 pix$^{-1}$ (MIPS). Processed data are given in units of MJy\,sr$^{-1}$, and for the photometric analysis we transformed them to $\mu$Jy after multiplication by factors of 8.4616 (IRAC) and 141.08 (MIPS). 

We measured aperture photometry for each object using the task PHOT within the IRAF\footnote{IRAF is distributed by the National Optical Astronomy Observatories, which are operated by the Association of Universities for Research in Astronomy, Inc., under cooperative agreement with the National Science Foundation.} environment. For IRAC, we used an aperture radius of 4 and 10 pix and inner and outer radii of 24 and 40 pix for the sky annulus. For MIPS, we adopted radii of 1.22, 7, and 12 pix for the aperture and the inner and outer boundaries of the sky annulus, respectively. We applied the appropriate aperture corrections for each band to obtain the final fluxes that were converted into magnitudes using the following zero-point fluxes (taken from the IRAC and MIPS Data Handbooks): 280.9 Jy ([3.6]), 179.7 Jy ([4.5]), 115.0 Jy ([5.8]), 64.13 Jy ([8.9]), and 7.14 Jy ([24]). Our photometry is presented in Table~\ref{spitzer}; error bars take into account small variations in the sky contribution, and do not include the uncertainties in the calibrations of IRAC and MIPS. 

The L0 dwarf J0241$-$03 is undetected in the MIPS 24\,$\mu$m image; in Table~\ref{spitzer} we provide an upper limit on its brightness based on the magnitude of a virtual object with a flux peak of four times the sky background root-mean-square ($rms$) around the expected position of the target. J0136$+$09, J0355$+$11, and J1022$+$58 were observed in the IRAC [4.5]-band on two different occasions separated by about 1.5, 1, and 5~yr, respectively. The two measurements are consistent within the quoted uncertainty, suggesting that variability at this wavelength is likely below 0.06~mag at a 3-$\sigma$ confidence level ($\sigma$ is the photometric error bar). 

\subsection{Mid-infrared {\sl WISE} photometry}
Four of our targets, J0241$-$03, J0355$+$11, and the two USco objects, also have mid-infrared photometry obtained with the Wide-field Infrared Survey Explorer ({\sl WISE}, Wright et al$.$ \cite{wright10}). Table~\ref{wise} provides the {\sl WISE} photometry for the following wavelengths: 3.3526, 4.6028, 11.5608, and 22.0883 $\mu$m. These data are extracted from the {\sl WISE} preliminary data release, which includes the first 105 days of {\sl WISE} survey observations between 2010 Jan 14 and 2010 Apr 29. We caution that the {\sl WISE} photometry is processed with initial calibrations and reduction algorithms. 

J0241$-$03 is not detected at the two longest wavelengths, and USco\,128 and J0355$+$11 have poor [22]-band photometry. The emission of USco\,128 is remarkable at 11.56 $\mu$m, thus confirming the mid-infrared flux excess reported by Jayawardhana et al$.$ \cite{jayawardhana03} and Scholz et al$.$ \cite{scholz07}. Using the {\sl WISE} data we do not observe significant flux excesses up to 4.6 $\mu$m in USco\,132, which contrasts with the results by Jayawardhana et al$.$ \cite{jayawardhana03}.

The {\sl Spitzer} [4.5] and {\sl WISE} [4.6] magnitudes are quite similar for the L dwarfs J0241$-$03 and J0355$+$11, while the {\sl Spitzer} [3.6] magnitude appears to be 0.25--0.4 mag brighter than the {\sl WISE} [3.4]-band data. This is likely because the {\sl WISE} [3.4] filter is broader than the {\sl Spitzer} one, it extends towards the blue wavelengths covering regions of the L dwarf spectra deeply absorbed by water vapor not explored by the {\sl Spitzer} filter. On the contrary, the widths of the {\sl WISE} [4.6] and {\sl Spitzer} [4.5] bands are more alike, and although the {\sl WISE} filter is slightly shifted to redder wavelengths, there is no strong molecular absorption at these frequencies in the spectra of M and L dwarfs. Wright et al$.$ \cite{wright11} also noted that the color term between {\sl Spitzer} [4.5]  and {\sl WISE} [4.6] is rather small ($[4.5]-[4.6] = 0.054$ mag) and seems to have no trend with spectral type for the T dwarfs. Nevertheless, as stated in Patten et al$.$ \cite{patten06} we caution that color terms may be present when transforming fluxes measured in approximately similar filters from two different systems, particularly if the sources have large flux variations across the filter bandpasses.

\subsection{Spectral energy distributions}
\begin{figure}
\plotone{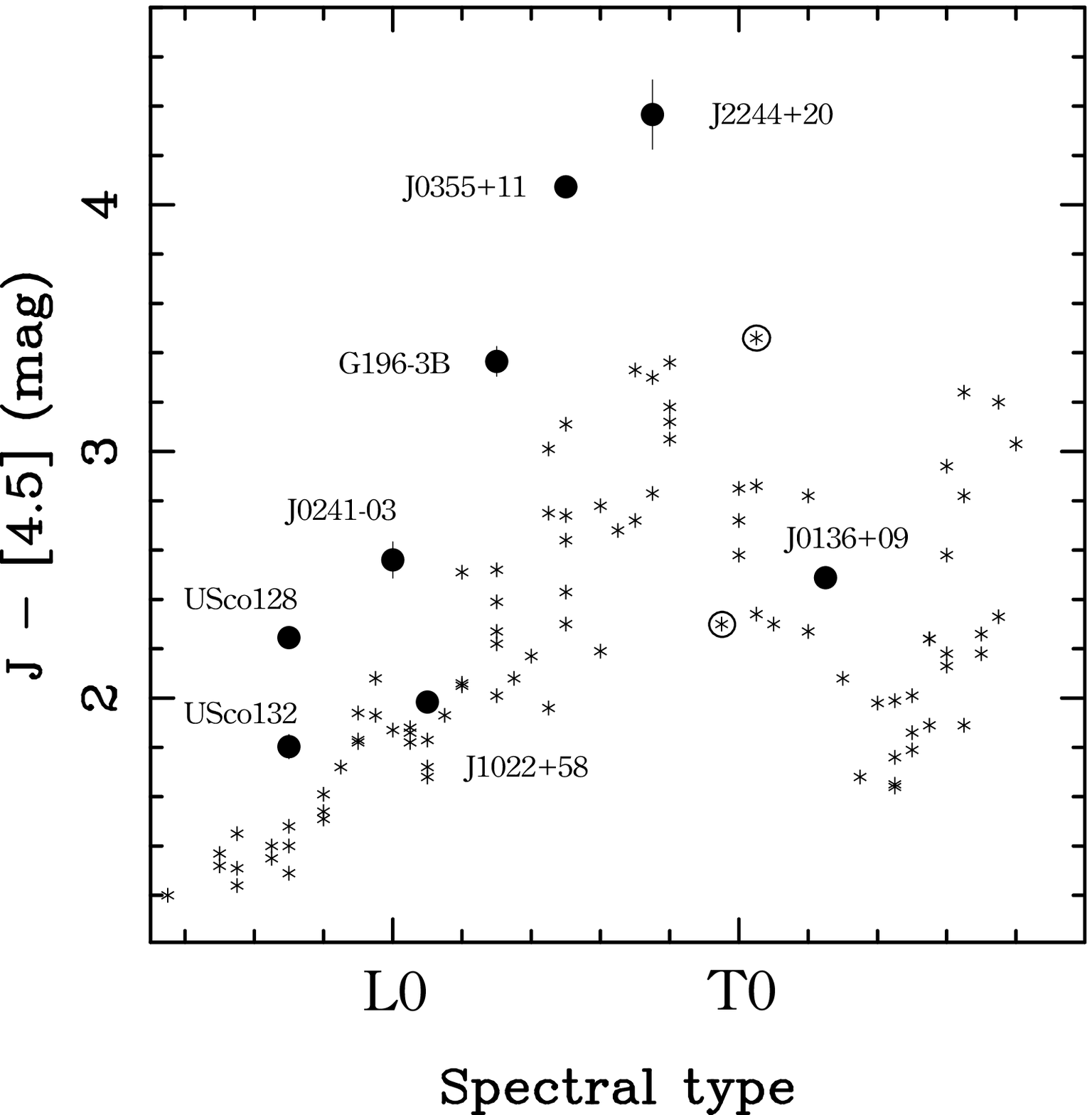}
\plotone{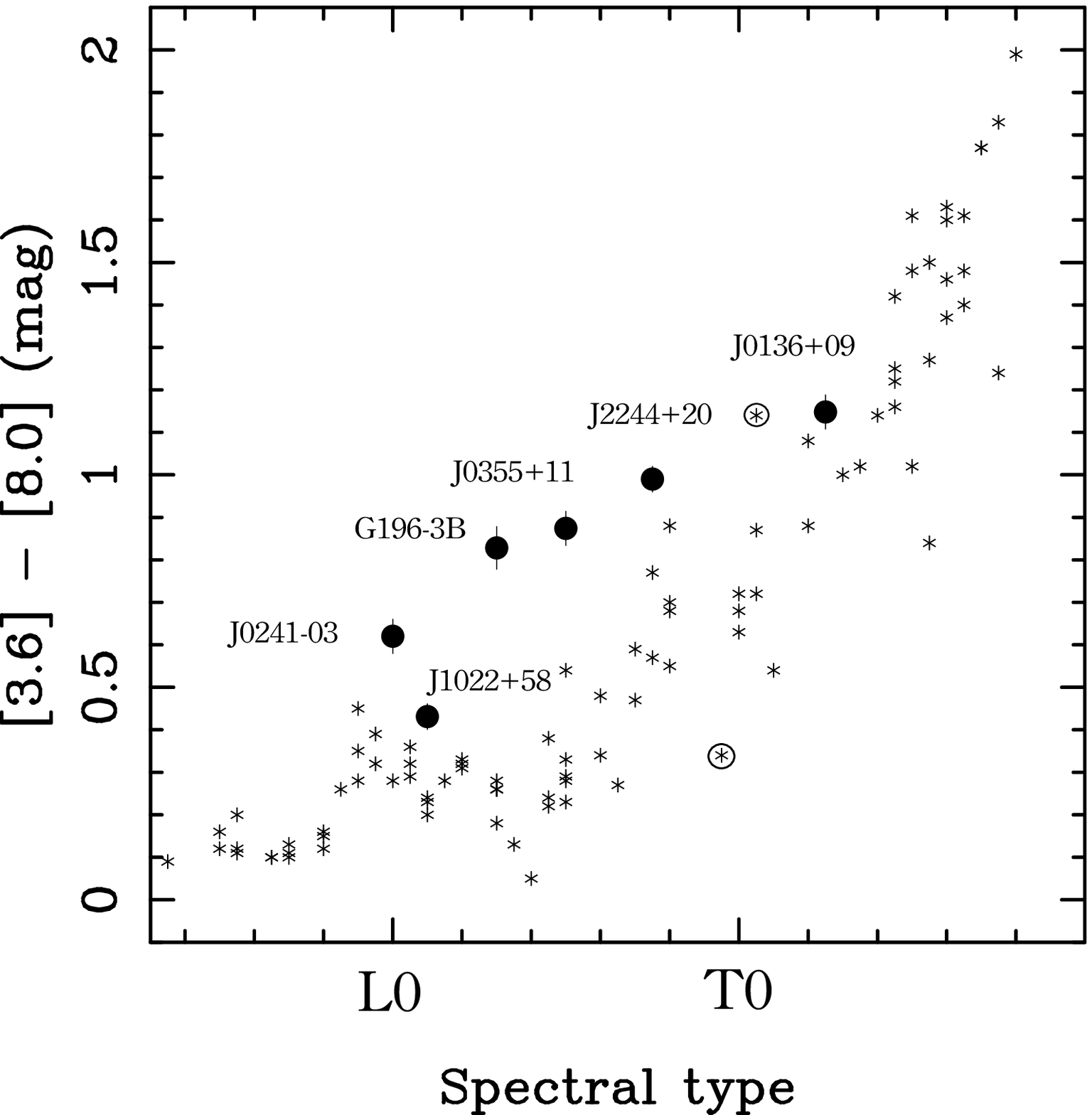}
\caption{Infrared colors of our science targets (labelled solid dots) as a function of spectral type. Field dwarfs of similar typing from Patten et al$.$ \cite{patten06} and Leggett et al$.$ \cite{leggett07} are plotted as asterisks. The reddish nature of the L-type targets is noticeable particularly in the upper panel. The uncertainty in spectral classification is half a subtype, except for J2244$+$20, which is of 1 subtype. Among Leggett et al.'s \cite{leggett07} sources (asterisks) there are one L9.5 dwarf with blue near-infrared colors and one T0.5 dwarf with red near-infrared colors (circled asterisks). \label{color}}
\end{figure}

In Fig.~\ref{color} we illustrate the $J-[4.5]$ and $[3.6]-[8.0]$ colors of our targets in comparison with the spectral sequence delineated by field dwarfs of ``normal" colors taken from Patten et al$.$ \cite{patten06} and Leggett et al. \cite{leggett07}. For the USco objects we use the {\sl WISE} [4.6] data by assuming that these magnitudes would be similar to the {\sl Spitzer} [4.5] photometry as it is the case for the L dwarfs. The reddish behavior of many of our targets becomes apparent in Fig.~\ref{color}. We quantify flux excesses in the $[4.5]$ band in Section~\ref{fexcess}. 

\begin{figure}
\plotone{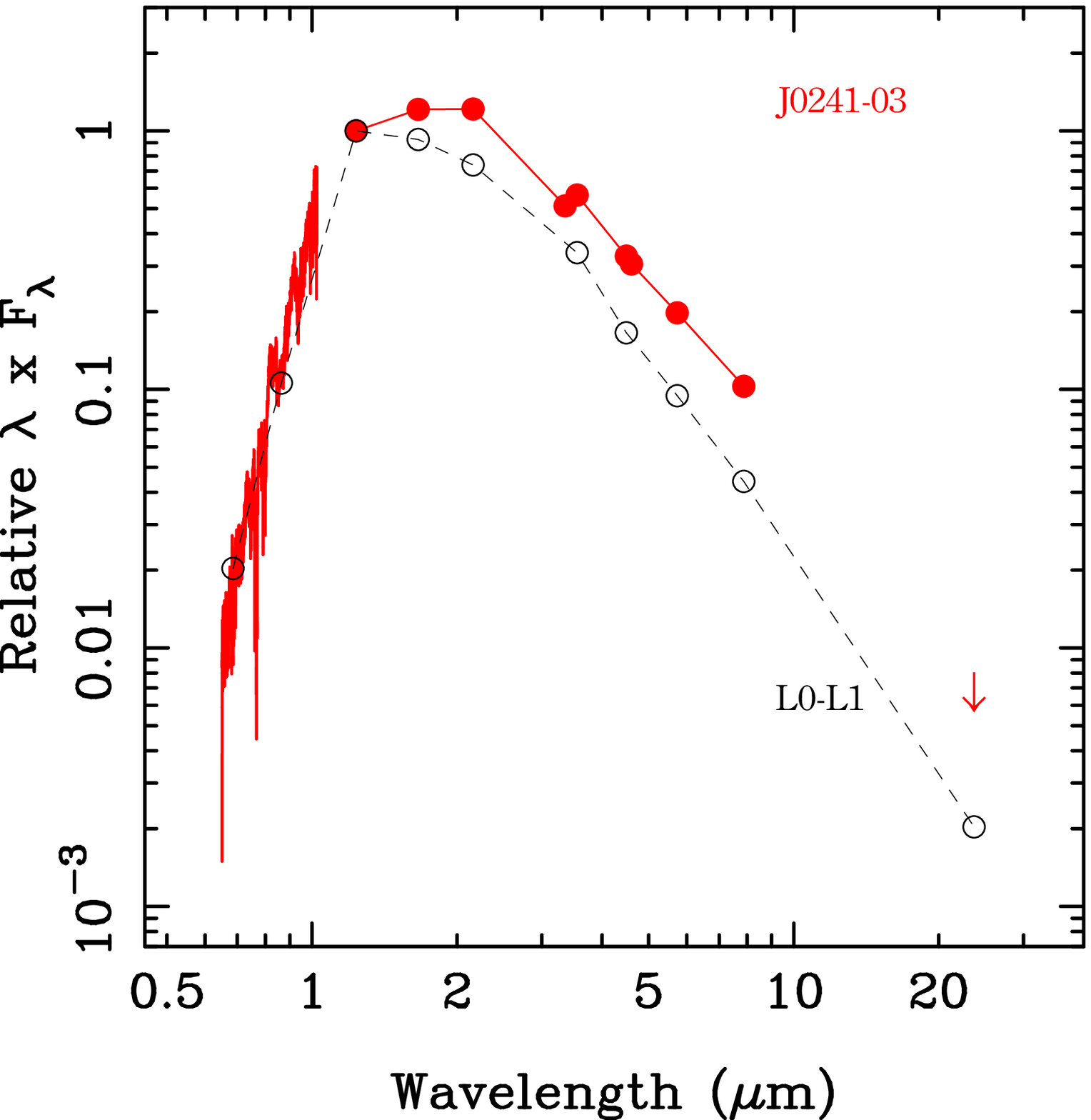}
\plotone{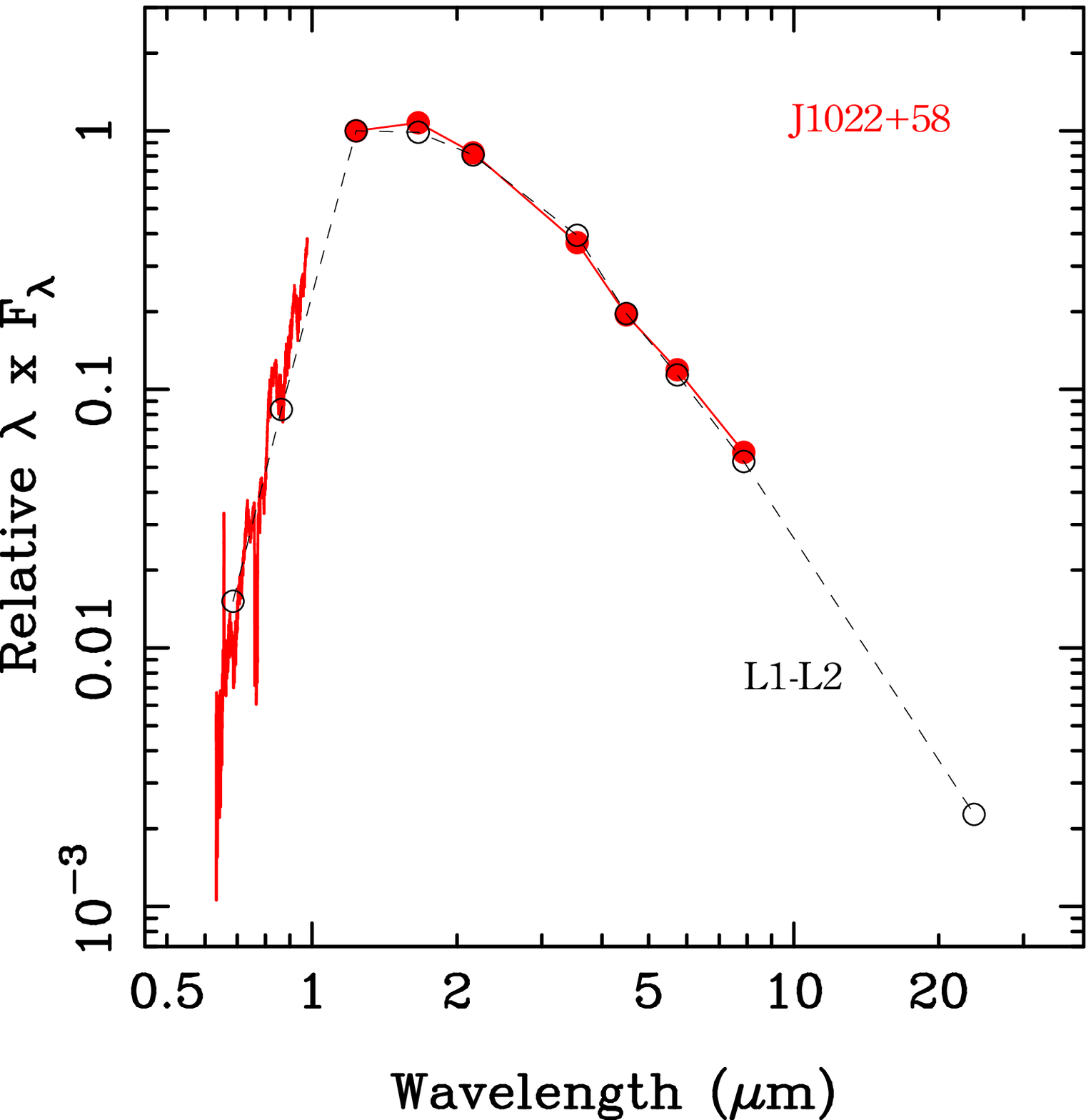}
\caption{Photometric spectral energy distributions (SEDs, red filled dots and solid lines) of J0241$-$03 (L0) and J1022$+$58 (L1). The SEDs include the optical spectra from Cruz et al$.$ \cite{cruz09}, normalized to the $R$ and $I$-bands, and the 2MASS, {\sl Spitzer} and {\sl WISE} photometry (see text). The 24-$\mu$m upper limit magnitude of J0241$-$03 is plotted as an arrow. The photometric 1-$\sigma$ uncertainty is smaller than the symbol size. For comparison purposes, overplotted are the average SEDs of field L0--L1 and L1--L2 dwarfs (black open circles and dashed lines).  All SEDs are normalized to unity  at the $J$-band (1.235~$\mu$m).  \label{seds}}
\end{figure}

All available broad-band photometry and low-resolution spectroscopy were combined to create visible and infrared spectral energy distributions (SEDs) for two objects in our sample: one with red infrared colors (J0241$-$03, L0) and another one with no obvious color deviations (J1022$+$58, L1). Observed magnitudes were converted to monochromatic fluxes at the central wavelength of each filter using the zero point fluxes provided in the literature for 2MASS and those mentioned in previous sections for the {\sl Spitzer} instruments. The zero point fluxes of the {\sl WISE} data were taken from Wright et al$.$ \cite{wright10}. The computed SEDs from 0.65 to 24~$\mu$m are provided in Fig.~\ref{seds}. For comparison with J0241$-$03 and J1022$+$58, we overplotted the average photometric SEDs of field dwarfs of similar spectral classification (L0--L1 and L1--L2) and no infrared flux excesses. We compiled the required photometry $R$ through $[24]$ from Liebert \& Gizis \cite{liebert06} (optical data), the 2MASS catalog (near-infrared), Patten et al$.$ \cite{patten06} (3--8 $\mu$m), and Zapatero Osorio et al$.$ \cite{osorio10}Ê($[24]$). From Fig.~\ref{seds}, the SED of J1022$+$58 clearly indicates its photospheric origin since there is no obvious mid-infrared flux excess up to 8 $\mu$m. In contrast, the SED of J0241$-$03 appears overluminous longwards of the $H$-band when compared to the average SED of L0--L1 dwarfs. G\,196--3\,B (L3) has a similar property as it is shown in Fig.~3 of Zapatero Osorio et al$.$ \cite{osorio10}. The polarimetric observations may shed new light on the understanding of the origin of this feature.

\subsection{Linear polarimetric observations}
Linear polarimetric images were acquired for our sample targets with the Long-slit Intermediate Resolution Infrared Spectrograph (LIRIS, Manchado et al$.$ \cite{manchado04}) mounted at the Cassegrain focus of the William Herschel Telescope (WHT) during the nights 2004 Oct 27, 2006 Mar 22, 2010 Oct 25, and 2011 Apr 20. The 2004 campaign was carried out as part of the LIRIS instrument commissioning on the WHT. LIRIS is equipped with a 1024\,$\times$\,1024 pix$^2$ Hawaii detector covering the wavelength range 0.8--2.5 $\mu$m. The pixel projection onto the sky is 0\farcs25 yielding a field of view of 4\farcm27\,$\times$\,4\farcm27. Polarimetric observations can be performed by using a Wedged double Wollaston device (Oliva \cite{oliva97}), consisting in a combination of two Wollaston prisms that deliver four simultaneous images of the polarized flux at angles 0 and 90, 45 and 135 deg. An aperture mask 4\arcmin\,$\times$\,1\arcmin~in size is used to prevent overlapping effects between the different polarization vector images. 

Our data were acquired using the $J$- and $H$-band filters and (in some occasions) two different angles of the rotator of the telescope. The latter is convenient to derive more accurate polarimetric measurements (Alves et al$.$ \cite{alves11}), since flat-field effects and optical path differences are reduced. The central wavelengths and passbands of the filters are 1.25 and 0.16~$\mu$m ($J$), and 1.64 and 0.29~$\mu$m ($H$). All program objects (except J2244$+$20) were observed in the $J$-band; one target (USco\,128) was observed in the two $JH$-bands, and J2244$+$20 was imaged only in the $H$-band. The Earth atmospheric conditions during the observations were photometric (good transparency) with mean seeing values of 0\farcs7--0\farcs8 in 2004 and 2006, 3\farcs0 in 2010, and 2\farcs0 in 2011. The log of the observations is shown in Table~\ref{log}, where we provide the list of targets (science and calibration sources), their $J$-band magnitudes and spectral types, Universal Time observing dates, exposure times (integration times were scaled according to the seeing and target brightness), angles of the telescope rotator, and the airmass range during data acquisition. The observing strategy consisted of acquiring frames following a 3-point (2004), a 5-point (2006),  and a 9-point (2010, 2011) dither pattern for a proper subtraction of the sky background contribution. The dither pattern cycle was repeated a few times for each target to increase the signal-to-noise ratio of the final measurements. 

For the calibration of the polarimetric measurements, observations of zero polarized standard stars (HD\,18803, Feige\,110, GD\,319, SA\,29$-$130, and HD\,14069, Schmidt et al$.$ \cite{schmidt92}; Clayton \& Martin \cite{clayton81}) and polarized standards (HD\,38563c, HRW\,24, Elia\,2$-$25, and HD\,283855, Whittet et al$.$ \cite{whittet92}) were carried out during the four observing campaigns. The white dwarfs Feige\,110 and SA\,29$-$130 are not included in the aforementioned references. We chose them as zero-polarized sources because non-magnetic white dwarfs in principle lack any intrinsic linear polarization. Both program and calibration objects were acquired at the same spot within the detector, which is very close to the optical axis of the telescope/instrument system.

We sliced individual frames into four smaller frames corresponding to each of the polarimetry vectors prior to raw data reduction. Each polarimetry vector was then processed separately. Frames were dark-current subtracted and divided by a   image taken with the polarimetric optics in the 2004, 2010, and 2011 campaigns. The 2006 data were not corrected for flat-field because there was an insufficient number of calibration images acquired with the polarimetric optics. Earth atmosphere background contribution was removed from all individual images, which were finally registered and stacked together to produce deep data. 

\subsection{Polarimetric analysis}
To derive the normalized Stokes parameters $q$ and $u$ along with the polarization degree $P$ and the angle of the polarization vibration $\Theta$, we employed the ``direct measurement" method (Eqs$.$ 1 and 2) for the data collected with the telescope rotator angle of $0\deg$, and the ``flux ratio" calculation method (Eqs$.$ 3 and 4) for the data taken with telescope rotator angles of 0 and $90\deg$. These methods are associated to the following mathematical equations:
\begin{equation}
q = \frac{i_{0,0} - i_{90,0}\,t_{0}/t_{90}}{i_{0,0} + i_{90,0}\,t_{0}/t_{90}}  
\end{equation}
\begin{equation}
u = \frac{i_{45,0} - i_{135,0}\,t_{45}/t_{135}}{i_{45,0} + i_{135,0}\,t_{45}/t_{135}}  
\end{equation}
\begin{equation}
R^2_q = \frac{i_{0,0} / i_{90,0}}{i_{0,90} / i_{90,90}} ; ~~~~
q = \frac{R_q - 1}{R_q + 1}
\end{equation}
\begin{equation}
R^2_u = \frac{i_{45,0} / i_{135,0}}{i_{45,90} / i_{135,90}} ; ~~~~
u = \frac{R_u - 1}{R_u + 1}
\end{equation}
\begin{equation}
P = \sqrt{(q-q_{\rm ins})^2 + (u-u_{\rm ins})^2}  \label{eqpol}
\end{equation}
\begin{equation}
\Theta = 0.5~{\rm atan}(u / q) - \Theta_0  \label{angle}
\end{equation}
where $i_{\rm vector,rot}$ stands for the flux per time unit corresponding to a given polarization vector (0, 45, 90, and 135 deg) and observed at a particular position of the telescope rotator (0 and 90 deg, see Table~\ref{log}); the factors $t_{0}/t_{90}$ and $t_{45}/t_{135}$ represent the relative transmission efficiency of each Wollaston; $q_{\rm ins}$ and $u_{\rm ins}$ stand for polarization induced by the telescope$+$instrument configuration and $\Theta_0$ is the zero-point offset in the angle of the polarization vibration, both of which likely depend on wavelength. The $\Theta$ angle is usually given in the interval 0--180 deg. The quantities $R_q$ and $R_u$ correct for possible flat flaws and optical path differences. 

We obtained the fluxes for each vector by determining circular aperture photometry using the tasks IMEXAMINE and PHOT within IRAF. In Tables~\ref{pol1} and~\ref{pol2} we list the radius of aperture annulus employed for each target and filter, and the full-width half-maximum (FWHM) measured on the reduced data. Photometric apertures ranged from 1.5 to 6 times the FWHM, the median aperture size was around 5.5 times larger than the FWHM. Both IRAF tasks performed a second sky background subtraction by computing the sky level in a ring outside the photometric aperture annulus; we found it necessary in order to remove some residuals from the previous data reduction process. G\,196--3\,B is well resolved spatially from its primary star in the 2006 data, therefore we do not expect contamination in our linear polarization measurement. 

The relative transmission factors of the Wollaston prisms were determined by assuming that non-polarized standard stars have null Stokes parameters $q$ and $u$. In this process associated to the direct measurement method, we would be correcting for any small instrumental polarization, and $q_{\rm ins}$ and $u_{\rm ins}$ would become zero in Eq$.$ \ref{eqpol}. The derived values of the transmission factors for each filter and observing night, and the employed standard stars are provided in Table~\ref{factors}. The various measurements for the 0--90 deg Wollaston are consistent within the quoted uncertainties; a larger difference is observed for the 45--135 deg Wollaston. In the flux ratio method, the contribution of the transmission factors is cancelled out; however, some instrumental polarization may remain. This was checked with the non-polarized standard stars and the observations acquired with telescope rotator angles of 0 and $90\deg$. We measured a small instrumental polarization of $q_{\rm ins} = +0.163 \pm 0.050 \%$ and $u_{\rm ins} = -0.168 \pm 0.050\%$ for the $J$-band in 2010 Oct 25, and no significant instrumental polarization in the $H$-band. The observations of polarized standard stars allowed us to check for the efficiency of the instrument. We found that our measurements of $P$ (after correction from instrumental polarization) and $\Theta$ are fully consistent with the literature values within 1-$\sigma$ the associated error bars. There is a zero-point correction to be applied to the position angle of the polarization vibration, which is $\Theta_0$\,=\,$+$4.46\,$\pm$\,1.5~deg for the $J$-band. The instrumental contribution was  removed from our data and the angle offset was taken into account following Eqs.~\ref{eqpol} and~\ref{angle} to produce the final polarimetric measurements given in Tables~\ref{pol1} (standard stars) and~\ref{pol2} (science targets). In column 11 of Table~\ref{pol1} we provide published polarimetric data for the standard stars permitting a proper comparison with our determinations.

For all objects we provide $P$ obtained with the direct measurement method in column 8 of Tables~\ref{pol1} and~\ref{pol2}. When applicable, columns 6, 7, 9, and 10 of both Tables list the measurements derived from the flux ratio method (telescope rotator angles of 0 and 90 deg). The uncertainty in the polarization degree is obtained as the quadratic combination of the instrumental polarization error (0.05--0.10\%) and the dispersion of the source polarization degree derived by using different photometric aperture radii. Uncertainties in polarization angle are not well defined for the unpolarized sources, and thus are not listed in Tables~\ref{pol1} and~\ref{pol2}. The linear polarimetry degrees are quoted as measured, i.e., without any correction for instrumental efficiency loss or correction for the statistical bias which affects polarimetry at small values of polarimetric signal-to-noise ratio (e.g., Simmons \& Stewart \cite{simmons85}). We note that the error bars of the polarimetric data derived from two telescope rotator angles are a factor of two smaller than those of data collected at 0 deg; this result is in agreement with the findings by Alves et al$.$ \cite{alves11}. Along with their spectral types, polarization degrees, position angles of the plane of vibration, and their associated uncertainties, columns 11 and 12 of Table~\ref{pol2} list the science targets spectroscopic rotational velocities and rotational periods when known in the literature (see Section~\ref{targets}). 

\section{Discussion}
To select the likely polarized sources amongst our targets, we applied the following criterion widely used in the literature (e.g., M\'enard et al$.$ \cite{menard02}; Zapatero Osorio et al$.$ \cite{osorio05}; Goldman et al$.$ \cite{goldman09}): the derived degree of polarization must exceed three times the associated uncertainty ($P/\sigma_P \ge 3$). On the assumption of a Gaussian distribution of the measurements within their error bars, such a criterion sets the confidence of positive detections at the level of 99\%. In the sample of eight targets, only one shows a clear detection: J0241$-$03  ($P/\sigma_P \sim 10$).  Since the polarization degree $P$ is a positive quantity, it would be overestimated by the error in $P$ for unpolarized sources. The debiased polarization degree of the non-polarized targets in Table~\ref{pol2} would become close to 0\,\%~by applying the equation $P' = \sqrt{P^2 - \sigma_P^2}$ (Wardle \& Kronberg \cite{wardle74}). Therefore, from our data we can impose the following 3-$\sigma_P$ upper limits on the near-infrared linear polarization degree: $\le$0.9\,\%~for J2244$+$20 ($H$), J1022$+$58 ($J$), and J0136$+$09 ($J$), and $\le$1.8\,\%~for J0355$+$11, G\,196--3\,B, USco\,128, and USco\,132 ($J$). The latter upper limit is not very restrictive, implying that more accurate data are required to study the presence of any significant linear polarization in these dwarfs. Next, we shall focus the discussion on the four targets with the best polarimetric constraints.

\subsection{Infrared color excess and polarization \label{fexcess}}
\begin{figure}
\plotone{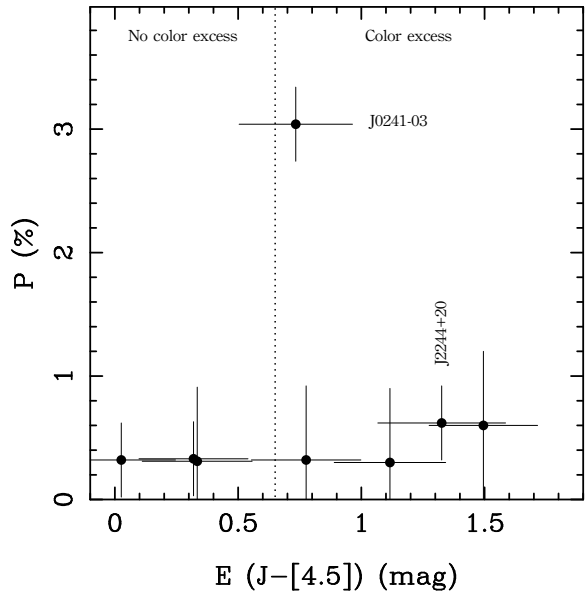}
\caption{$J$ (all targets, except J2244$+$20) and $H$ (J2244$+$20) linear polarization degree as a function of $J-[4.5]$ color excess. The source with the largest linear polarization intensity is J0241$-$03 (L0). We plot the average linear polarization  value for G\,196$-$3\,B. The vertical dotted line separates late-M and L-type objects with clear $J-[4.5]$ color excesses from those with no excesses at all. This line is defined as three times the dispersion of the average colors of field dwarfs. There is no obvious correlation between linear polarization degree in the near-infrared and infrared color excess. \label{excess}}
\end{figure}

Figure~\ref{excess} displays our measured linear polarization degree against $J-[4.5]$ color excess for our sample. The color excess $E(J-[4.5])$, given in the last column of Table~\ref{pol2}, is computed as the difference between the target's measured color and the average color of field dwarfs of the same spectral classification. We derived the mean $J-[4.5]$ index for the field using the data published in Patten et al$.$ \cite{patten06} and Leggett et al$.$ \cite{leggett07} and fitting third- and second-order polynomials to the spectral ranges M5--L9 and T0--T4, respectively. The fits have $rms$ of 0.218 (M5--L9) and 0.237 mag (T0--T4). Five targets (J2244$+$20, G\,196$-$3\,B, USCo\,128, J0241$-$03, and J0355$+$11) show significant color excesses at the level of 3--7 times the $rms$ of the polynomial fits. All except J2244$+$20 are thought to have low gravity atmospheres and an age younger than 300 Myr. Out of the four young sources, only one (J0241$-$03) is linearly polarized in the $J$-band at a level of $\ge$1.8\%~(3-$\sigma_P$ confidence), i.e., 25$\pm$25\,\%. 

Among the three objects, J0136$+$09, J1022$+$58 and USCo\,132, with no obvious $J-[4.5]$ color excess, none seems to be convincingly confirmed young sources with an age below 300 Myr. J0136$+$09 is a field T dwarf with quite ``standard" colors, USCo\,132 is not spectroscopically confirmed as a member of the Upper Scorpius region (Muzerolle et al$.$ \cite{muzerolle03}), and the high Galactic velocity of J1022$+$58 appears to indicate that it is a probable member of the Galactic thick disk (Blake et al$.$ \cite{blake10}), although such a high velocity might also result from an ejection process. All three sources are unpolarized in the near-infrared.

As illustrated in Fig.~\ref{excess}, there is no obvious correlation between the linear polarization index and the infrared color excess, except for the fact that the one positive detection happens to be a source with infrared flux excesses. This is likely due to the small number of targets and the role played by other parameters like age, metallicity, and rotation, all of which are critical to determine the amount of dust and the oblateness of the atmospheres. 

\subsection{The polarized case}
Sengupta \& Marley \cite{sengupta10} have recently modeled the expected intrinsic linear polarization intensity of ultracool dwarfs with effective temperatures between 1300 and 2400 K. This basically covers the spectral types of our L-type and early-T sample. The authors have considered that the light scattering by atmospheric dust particles is responsible for the polarization, and have incorporated two possible scenarios in their theory: polarization arising from inhomogeneous distributions of dust in the atmosphere (e.g., dust clouds) and from an oblate shape induced by rapid rotation. The conclusions based on the comparison of the models to $I$-band observations are that the second scenario is more likely and that low gravity atmospheres favor higher polarization degree than high gravities because the non-sphericity shape is more pronounced at low gravities. From Fig.~3 of Sengupta \& Marley \cite{sengupta10}, the predicted linear polarization degree is below 0.3\,\%~for the near-infrared $JHK$-bands and for atmospheres with surface gravity log\,$g$\,=\,4.5 (cm\,s$^{-2}$).  Furthermore, the linear polarization seems to be maximum at temperatures around 1800~K, i.e., spectral type $\sim$L2. Unfortunately, the precision of our polarimetric observations is not good enough to detect such low values of polarization. 

The only positive detection in our work, J0241$-$03 (L0), has $P = 3.04\pm0.30$\,\%, which clearly deviates from the predictions by Sengupta \& Marley \cite{sengupta10}. A very rapid rotation would contribute to increase the polarization intensity. The models depicted by these authors are already computed for a relatively high rotation ($v$\,sin\,$i$ = 48 km\,s$^{-1}$); to our knowledge, the fastest projected rotational velocity measured in large samples of late-M and L-type sources is around 70 km\,s$^{-1}$ (Reiners \& Basri \cite{reiners10}, and references therein). The rotational velocity of J0241$-$03 is unknown, but it likely lies in the range 0--70 km\,s$^{-1}$. This suggests that the origin of the observed light polarization (at least great part of it) is not intrinsic to the object's dusty atmosphere. A local circum(sub)stellar environment, a local molecular cloud, and foreground interstellar clouds may also cause detectable polarization. To the best of our knowledge, J0241$-$03 is not a member of any known star-forming region obscured by a molecular cloud, but a free-floating dwarf in the solar neighborhood. The optical study by Tamburini et al$.$ \cite{tamburini02}, based on more than 1000 stars, indicates that the contribution to the polarization by the interstellar medium becomes effective only after $\sim$70 pc. J0241$-$03 is probably located at a distance of 66$\pm$8 pc (Cruz et al$.$ \cite{cruz09}); therefore, we do not expect interstellar polarization to contribute significantly to our measurement. 

This leaves us with one likely explanation: the L0 dwarf is the host of a surrounding dusty envelope or disk, a fact consistent with the supposedly young age of the dwarf. Protoplanetary disks (both primordial and transitional or debris disks) are commonly detected around brown dwarfs in various young star-forming regions through observations of mid-infrared flux excesses (e.g., Natta et al$.$ \cite{natta02}; Luhman et al$.$ \cite{luhman05}; Allers et al$.$ \cite{allers06}; Caballero et al$.$ \cite{caballero07}). The linear polarization degree of J0241$-$03 is consistent with the amounts (from $\sim$1\%~up to 25\%) of intrinsic polarization found at near-infrared wavelengths for other brown dwarfs in rather young, extinguished regions like M42 (Kusakabe et al$.$ \cite{kusakabe08}), NGC\,2024 (Kandori et al$.$ \cite{kandori07}), Taurus, $\rho$\,Ophiuci and Chamaeleon (Hashimoto et al$.$ \cite{hashimoto09}). It is well established that the existence of disks is a function of time: the disk frequency is higher at the youngest ages, yet debris disks can be seen up to ages of about a hundred Myr, e.g., 32\%~and 15\%~of the solar-mass stars of Blanco~1 and the Pleiades, both clusters with a similar age of 120 Myr, show flux excesses at 24 $\mu$m (Stauffer et al$.$ \cite{stauffer10}; Gorlova et al$.$ \cite{gorlova06}). On the contrary, less than 12\%~of the FGK stars of the Hyades ($\sim$ 600 Myr) show evidence for debris disks (Cieza et al$.$ \cite{cieza08}). Therefore, our observations fully support the youth of J0241$-$03 since detection of high degrees of polarization is usually regarded as evidence for dusty disks. The polarimetric measurement also hints at an asymmetric distribution of the grains responsible for the polarization within the disk, otherwise net polarization should have been null. This asymmetry appears standard in disks of brown dwarfs since 16 out of 34 (i.e., 47\%) young, disk-bearing brown dwarfs of Taurus, $\rho$\,Ophiuci and Chamaeleon show detectable near-infrared polarization (Hashimoto et al$.$ \cite{hashimoto09}). Given the proximity of J0241$-$03, this object becomes an interesting source for high-resolution imaging and direct detection of the disk (e.g., using ALMA or other instruments providing high-spatial resolution images). Despite the low number of near-infrared observations presented in this work, the derived frequency (25\,$\pm$\,25\%) of polarized ($P \ge 1.8\%$) young field brown dwarfs is consistent with the frequency found by Hashimoto et al$.$ \cite{hashimoto09}.

\subsection{Photometric variability and polarization}
The two sources with reported photometric variability attributed to heterogeneous distribution of atmospheric condensate clouds, J2244$+$20 and J0136$+$09, do not display significant linear polarization. We note that the polarimetric data are not contemporaneous with the photometric monitoring carried out by Morales Calder\'on et al$.$ \cite{morales06} and Artigau et al$.$ \cite{artigau09}. Our observations neither confirm nor reject the presence of structures of condensates in the atmospheres of the L7.5 and T2.5 dwarfs. 

According to Sengupta \& Marley \cite{sengupta10} and Sengupta \& Kwok \cite{sengupta05}, polarization induced by light scattering of atmospheric dusty grains is expected to be larger at optical wavelengths than at the near-infrared for L-type and early-T objects; this is mainly due to the predicted size and development of the grains and the amounts of condensates. J2244$+$20 was reported to be polarized in the $I$-band (0.85 $\mu$m) at a level of $P$ = 2.4\,\%~(5.2 $\sigma_P$  significance) by Zapatero Osorio et al$.$ \cite{osorio06}, although it was the faintest target in their sample. More recently, Goldman et al$.$ \cite{goldman09} did not detect optical polarization above the 0.76\,\%~level (2-$\sigma$). Either the errors were underestimated in the 2006 optical polarimetric measurement (quite likely) and J2244$+$20 is unpolarized, or this object may have significant linear polarimetry variability. The polarimetric non-detection of the T2.5 dwarf J0136$+$09 is consistent with the predictions by Sengupta \& Marley \cite{sengupta09}; based on state-of-the-art knowledge of T dwarf atmospheres, these authors concluded that, from the theoretical point of view, linear polarization of cloudless T dwarfs by atomic and molecular scattering may not be detectable at the red optical and near-infrared wavelengths, while dusty T-dwarfs may show a small amount of polarization.

\section{Conclusions and final remarks}
We carried out polarimetric imaging observations of eight ultracool dwarfs (most likely brown dwarfs) with spectral types in the interval M7--T2.5 using the $J$- and $H$-band filters and the LIRIS instrument on the WHT. The target list is formed by objects with detected photometric variability due to dust clouds, very red $J-K_s$ colors (as compared to other field dwarfs of similar classification), and low gravity atmospheres. We also obtained {\sl Spizter} IRAC and MIPS photometry (from the public archive) for five objects in the sample; in addition, four sources are included in the recent release of the {\sl WISE} catalog. This has allowed us to study the $J-[4.5]$ color of all targets in comparison to the mean color of field dwarfs with related spectral types, finding that five sources show clear $J-[4.5]$ color excesses, four of which are spectroscopically confirmed young objects with likely ages below 300 Myr. From the LIRIS data, we can impose 99\%-confidence level upper limits of $P \le 0.9$\%~on the near-infrared linear polarization degree for J2244$+$20, J0136$+$09, and J1022$+$58, and $P \le 1.8$\%~for G\,196$-$3\,B, J0355$+$11, and USco\,128 and USco\,132. We detected significant $J$-band linear polarization ($P = 3.04\pm0.30$) in the L0 dwarf J0241$-$03, which is one of the young brown dwarfs with notorious infrared color excesses: $E(J-[4.5]) = 0.73\pm0.23$ mag. The likely scenario to account for both the infrared flux excesses and the polarized light is the presence of a surrounding dusty envelope or disk with an asymmetric distribution of grains. The study of the linear polarization degree and polarization angle at different wavelengths may shed new light on the properties of the grains responsible for the observed polarization in J0241$-$03. The incidence of positive linear polarimetric detections with degree $P \ge 1.8$\%~in our sample of very red young sources ($\le$300 Myr)  is $25\pm25$\%.

\acknowledgments
This work is based in part on observations made with the William Herschel Telescope operated by the Isaac Newton Group on the island of La Palma at the Spanish Observatorio del Roque de los Muchachos of the Instituto de Astrof\'\i sica de Canarias (IAC), on observations made with the Spitzer Space Telescope, which is operated by the Jet Propulsion Laboratory, California Institute of Technology under a contract with the National Aeronautics and Space Administration (NASA), and on data products from the Wide-field Infrared Survey Explorer, which is a joint project of the University of California, Los Angeles, and the Jet Propulsion Laboratory/California Institute of Technology, funded by the NASA. This work is partly financed by the Spanish Ministry of Science through the projects AYA2010-21308-C03-02 and AYA2010-20535, and CSIC project 200950I010.

\clearpage


\clearpage

\begin{deluxetable}{lrcccccc} 
\tabletypesize{\scriptsize}
\tablecolumns{8} 
\tablecaption{Observing log\label{log}} 
\tablewidth{0pc} 
\tablehead{ 
\colhead{Object} & \colhead{$J$\tablenotemark{a}} & \colhead{SpT} & \colhead{Obs$.$ date} & \colhead{Filter} & \colhead{Rotator\tablenotemark{b}} & \colhead{Exposure time} & \colhead{Airmass} \\
\colhead{} & \colhead{(mag)} & \colhead{} & \colhead{} & \colhead{} & \colhead{(deg)} & \colhead{(s)} & \colhead{} 
}
\startdata 
2MASS\,J22443167$+$2043433    & 16.48   & L7.5 & 2004 Oct 28 & $H$ & 0, 90 & $4\times3\times2\times40, 4\times3\times2\times40$ & 1.05--1.02 \\
G\,196$-$3\,B                 & 14.83   & L3   & 2006 Mar 23 & $J$ & 0     & $3\times5\times60$                     & 1.27--1.34 \\
                              &         &      & 2011 Apr 20 & $J$ & 0, 90 & $2\times9\times120, 1\times9\times120$ & 1.10--1.13 \\
UScoCTIO\,128                 & 14.40   & M7   & 2006 Mar 23 & $J$ & 0     & $3\times5\times60$                     & 1.67--1.63 \\ 
                              &         &      & 2006 Mar 23 & $H$ & 0     & $3\times5\times60$                     & 1.63--1.68 \\
UScoCTIO\,132                 & 14.26   & M7   & 2006 Mar 23 & $J$ & 0     & $3\times5\times60$                     & 1.66--1.77 \\ 
2MASS\,J01365662$+$0933473    & 13.46   & T2.5 & 2010 Oct 25 & $J$ & 0, 90 & $3\times9\times120, 3\times9\times120$ & 1.16--1.06 \\
2MASS\,J02411151$-$0326587    & 15.80   & L0   & 2010 Oct 26 & $J$ & 0, 90 & $4\times9\times120, 4\times9\times120$ & 1.19--1.59 \\
2MASS\,J03552337$+$1133437    & 14.05   & L5   & 2010 Oct 26 & $J$ & 0     & $3\times9\times120$                    & 1.30--1.89 \\
2MASS\,J10224821$+$5825453    & 13.50   & L1   & 2011 Apr 20 & $J$ & 0, 90 & $1\times9\times120, 1\times9\times120$ & 1.15--1.54 \\
\tableline
HD\,38563c\tablenotemark{c}   &  9.73   & A0II & 2004 Oct 27 & $H$ & 0, 90 & $15\times10\times1, 15\times10\times1$ & 1.25--1.35 \\
                              &         &      & 2010 Oct 26 & $J$ & 0, 90 & $1\times9\times60, 1\times9\times60$   & 1.71--1.52 \\ 
HD\,283855\tablenotemark{c}   &  9.50   & A2   & 2004 Oct 28 & $H$ & 0, 90 & $9\times10\times3, 9\times10\times3$   & 1.00--1.01 \\
HD\,14069\tablenotemark{d}    &  8.53   & A0   & 2004 Oct 29 & $H$ & 0, 90 & $50\times3\times1, 50\times3\times1$   & 1.21--1.13 \\
HRW\,24\tablenotemark{c}      & 11.69   & A0V  & 2006 Mar 22 & $J$ & 0     & $2\times5\times5$                      & 1.74--1.90 \\ 
                              &         &      & 2006 Mar 22 & $H$ & 0     & $3\times5\times5$                      & 1.93--1.97 \\ 
GD\,319\tablenotemark{d}      & 11.54   & DA   & 2006 Mar 23 & $J$ & 0     & $3\times5\times2$                      & 1.14--1.13 \\
                              &         &      & 2006 Mar 23 & $H$ & 0     & $3\times5\times2$                      & 1.20--1.19 \\
Feige\,110\tablenotemark{d}   & 12.55   & DA   & 2010 Oct 25 & $J$ & 0     & $3\times9\times90$                     & 1.22--1.20 \\
HD\,18803\tablenotemark{d}    &  5.34   & G8V  & 2010 Oct 26 & $J$ & 0, 90 & $2\times9\times1, 2\times9\times1$     & 1.19--1.24 \\ 
SA\,29$-$130\tablenotemark{d} & 13.64   & DA   & 2011 Apr 20 & $J$ & 0, 90 & $1\times9\times120, 1\times9\times120$ & 1.04--1.03 \\
Elia\,2$-$25\tablenotemark{c} &  8.86   & B4   & 2011 Apr 20 & $J$ & 0, 90 & $2\times9\times10, 2\times9\times10$   & 3.69--3.14 \\
\enddata 
\tablenotetext{a}{2MASS $J$ band photometry (Skrutskie et al$.$ \cite{skrutskie06}).}
\tablenotetext{b}{Angle of the telescope rotator.}
\tablenotetext{c}{Polarized standard stars from Whittet et al$.$ \cite{whittet92}. }
\tablenotetext{d}{Zero-polarization standard stars are either white dwarfs or objects from Schmidt et al$.$ \cite{schmidt92} and Clayton \& Martin \cite{clayton81}.}
\end{deluxetable}

\begin{deluxetable}{lccccccc} 
\tabletypesize{\scriptsize}
\tablecolumns{8} 
\tablecaption{Spitzer infrared photometry\label{spitzer}} 
\tablewidth{0pc} 
\tablehead{ 
\colhead{Object} &  \colhead{[3.6]} & \colhead{[4.5]} & \colhead{[5.8]} & \colhead{[8.0]} & \colhead{[24]} & \colhead{AOR}  & \colhead{Observing date} \\
\colhead{(abridged)} & \colhead{(mag)} & \colhead{(mag)} & \colhead{(mag)} & \colhead{(mag)} & \colhead{(mag)} & \colhead{} & \colhead{} 
}
\startdata 
J2244$+$20    & 12.35$\pm$0.01\tablenotemark{a} & 12.11$\pm$0.01\tablenotemark{a} & 11.59$\pm$0.01\tablenotemark{a} & 11.59$\pm$0.02\tablenotemark{a} & 10.88$\pm$0.20 & 22139136\tablenotemark{b} & 2007 Jul 13\tablenotemark{b} \\
G\,196$-$3\,B & 11.66$\pm$0.02\tablenotemark{c} & 11.47$\pm$0.04\tablenotemark{c} & 11.10$\pm$0.06\tablenotemark{c} & 10.83$\pm$0.04\tablenotemark{c} & 10.55$\pm$0.10\tablenotemark{c} & \nodata & \nodata \\
USco\,128     & \nodata& \nodata & \nodata & \nodata & ~\,8.58$\pm$0.11\tablenotemark{d} & \nodata & \nodata \\ 
USco\,132     & \nodata& \nodata & \nodata & \nodata & \nodata & \nodata & \nodata \\
J0136$+$09    & 11.36$\pm$0.02 & 10.97$\pm$0.02 & 10.55$\pm$0.03 & 10.21$\pm$0.03 & ~\,9.48$\pm$0.15  & 21967360, 21911808\tablenotemark{e} & 2008 Feb 02, 2007 Aug 23\tablenotemark{e} \\
              &    \nodata          & 10.99$\pm$0.02 &  \nodata             & \nodata               &  \nodata                & 34699776 & 2009 Aug 22 \\
J0241$-$03    & 13.36$\pm$0.02\tablenotemark{f} & 13.24$\pm$0.02\tablenotemark{f} & 13.04$\pm$0.03\tablenotemark{f} & 12.77$\pm$0.03\tablenotemark{f} & $\ge$12.15  & 22137856\tablenotemark{b} & 2007 Aug 17\tablenotemark{b} \\
J0355$+$11    & 10.29$\pm$0.02 & ~\,9.98$\pm$0.02 & ~\,9.59$\pm$0.03 & ~\,9.41$\pm$0.03 & \nodata  & 25363712 & 2008 Sep 21 \\
              &    \nodata            & 10.00$\pm$0.02   &  \nodata                 &   \nodata               &  \nodata & 34707456 & 2009 Oct 06 \\
J1022$+$58    & 11.55$\pm$0.02 & 11.51$\pm$0.02 & 11.29$\pm$0.02 & 11.11$\pm$0.02 & \nodata & 10380800 & 2004 Nov 24 \\
              &    \nodata     & 11.50$\pm$0.02 &    \nodata     &    \nodata     & \nodata & 34721792 & 2009 Dec 17 \\
\enddata 
\tablecomments{We provide here our {\sl Spitzer} photometry unless it is indicated otherwise.}
\tablenotetext{a}{IRAC photometry from Leggett et al$.$ \cite{leggett07}.}
\tablenotetext{b}{Corresponding to MIPS ([24]) photometry.}
\tablenotetext{c}{IRAC and MIPS photometry from Zapatero Osorio et al$.$ \cite{osorio10}.}
\tablenotetext{d}{Observed magnitude obtained from the MIPS flux at 24\,$\mu$m given by Scholz et al$.$ \cite{scholz07}.}
\tablenotetext{e}{IRAC and MIPS photometry, relatively.}
\tablenotetext{f}{IRAC photometry from Luhman et al$.$ \cite{luhman09}.}
\end{deluxetable}

\begin{deluxetable}{lcccc} 
\tabletypesize{\scriptsize}
\tablecolumns{5} 
\tablecaption{WISE infrared photometry\label{wise}} 
\tablewidth{0pc} 
\tablehead{ 
\colhead{Object} &  \colhead{[3.4]\tablenotemark{a}} & \colhead{[4.6]\tablenotemark{a}} & \colhead{[12]\tablenotemark{a}} & \colhead{[22]\tablenotemark{a}} \\
\colhead{(abridged)} & \colhead{(mag)} & \colhead{(mag)} & \colhead{(mag)} & \colhead{(mag)}  
}
\startdata 
USco\,128  & 12.83$\pm$0.03 & 12.15$\pm$0.03 & 10.58$\pm$0.09 & 8.73$\pm$0.39\tablenotemark{b} \\
USco\,132  & 12.75$\pm$0.03 & 12.46$\pm$0.03 & 12.11$\pm$0.38\tablenotemark{b} & \nodata       \\
J0241$-$03 & 13.65$\pm$0.03 & 13.23$\pm$0.04 &  \nodata       & \nodata                        \\
J0355$+$11 & 10.55$\pm$0.02 &  9.95$\pm$0.02 &  9.20$\pm$0.04 & 8.86$\pm$0.53\tablenotemark{b} \\
\enddata 
\tablenotetext{a}{Isophotal wavelengths of the WISE filters are: 3.3526, 4.6028, 11.5608, and 22.0883 $\mu$m (Wright et al$.$ \cite{wright10}).}
\tablenotetext{b}{Marginal detection with signal-to-noise ratio of 2.8 (USco\,128), 2.9 (USco\,132), and 2.1 (J0355$+$11).}
\end{deluxetable}

\begin{deluxetable}{cccccl} 
\tabletypesize{\scriptsize}
\tablecolumns{6} 
\tablewidth{0pc} 
\tablecaption{Wollaston transmission factors\label{factors}} 
\tablehead{ 
\colhead{Date} & \colhead{Filter} & \colhead{$t_0/t_{90}$} & \colhead{$t_{45}/t_{135}$} & \colhead{Unpolarized star} & \colhead{Comments} \\
}
\startdata 
2004 Oct & $H$ & 1.0000$\pm$0.0052 & 1.0274$\pm$0.0054 & HD\,14069             & Flat-fielded images \\
2006 Mar & $J$ & 1.0039$\pm$0.0052 & 1.0540$\pm$0.0054 & GD\,319               & No flat-field correction\\
2006 Mar & $H$ & 0.9964$\pm$0.0052 & 1.0436$\pm$0.0054 & GD\,319               & No flat-field correction\\
2010 Oct & $J$ & 0.9966$\pm$0.0052 & 1.0370$\pm$0.0054 & Feige\,110, HD\,18803 & Flat-fielded images \\
\enddata 
\end{deluxetable}

\begin{deluxetable}{lcccrcccccc} 
\tabletypesize{\scriptsize}
\tablecolumns{11} 
\tablewidth{0pc} 
\tablecaption{Linear polarization measurements of standard stars\label{pol1}} 
\tablehead{ 
\colhead{Object} & \colhead{SpT} &  \colhead{Filter} & \colhead{FWHM} & \colhead{Ann.\tablenotemark{a}} & \colhead{$q$} & \colhead{$u$} & \colhead{$P$\tablenotemark{b}} & \colhead{$P$\tablenotemark{c}} & \colhead{$\Theta$} & \colhead{$P,\Theta$\tablenotemark{d}}  \\
\colhead{(abridged)} & \colhead{} & \colhead{} & \colhead{(\arcsec)} & \colhead{(\arcsec)} & \colhead{($\times 10^{-2})$} & \colhead{($\times 10^{-2})$} & \colhead{(\%)} & \colhead{(\%)} & \colhead{(deg)} & \colhead{(\%, deg)}\\
\colhead{(1)} & \colhead{(2)} & \colhead{(3)} & \colhead{(4)} & \colhead{(5)} & \colhead{(6)} & \colhead{(7)} & \colhead{(8)} & \colhead{(9)} & \colhead{(10)} & \colhead{(11)} 
}
\startdata 
HRW\,24       & A0V  &  $J$ & 0.6 &  3.6 & $-$2.08$\pm$0.42  & $+$0.59$\pm$0.42  &  2.16$\pm$0.60 & \nodata       &  82$\pm$10 & 2.10$\pm$0.05, 86$\pm$1 \\ 
              &      &  $H$ & 0.8 &  4.8 & $-$1.36$\pm$0.42  & $-$0.15$\pm$0.42  &  1.36$\pm$0.60 & \nodata       &  93$\pm$10 & 1.43$\pm$0.10, 88$\pm$1 \\ 
HD\,283855    & A2   &  $H$ & 1.8 & 10.0 & $+$0.09$\pm$0.21  & $+$1.46$\pm$0.21  &  1.69$\pm$0.60 & 1.46$\pm$0.30 &  43$\pm$10 & 1.67$\pm$0.06, 43$\pm$3 \\
HD\,38563c    & A0II &  $H$ & 0.7 &  3.8 & $-$3.53$\pm$0.21  & $+$1.80$\pm$0.21  &  4.28$\pm$0.60 & 3.96$\pm$0.30 &  76$\pm$10 & 3.68$\pm$0.10, 70$\pm$1 \\
              &      &  $J$ & 3.6 & 14.2 & $-$5.02$\pm$0.21  & $+$2.78$\pm$0.21  &  4.88$\pm$0.60 & 5.74$\pm$0.30 &  71$\pm$10\tablenotemark{e} & 6.03$\pm$0.10, 71$\pm$1\\ 
Elia\,2$-$25  & B4   &  $J$ & 2.9 & 17.3 & $+$3.75$\pm$0.21  & $+$5.27$\pm$0.21  &  5.78$\pm$0.60 & 6.47$\pm$0.30 &  23$\pm$10 & 6.46$\pm$0.05, 24$\pm$1 \\
HD\,14069     & A0   &  $H$ & 1.0 &  5.5 & $+$0.40$\pm$0.21  & $-$0.26$\pm$0.21  &  0.00$\pm$0.60 & 0.47$\pm$0.30 &  \nodata & \nodata \\
Feige\,110    & DA   &  $J$ & 2.8 & 13.0 & $+$0.06$\pm$0.42  & $-$0.20$\pm$0.42  &  0.20$\pm$0.60 & \nodata       &  \nodata & \nodata \\
HD\,18803     & G8V  &  $J$ & 2.7 & 13.0 & $-$0.00$\pm$0.21  & $+$0.01$\pm$0.21  &  0.35$\pm$0.60 & 0.01$\pm$0.30 &  \nodata & \nodata \\ 
GD\,319       & DA   &  $J$ & 0.6 &  3.6 & $+$0.00$\pm$0.42  & $+$0.00$\pm$0.42  &  0.00$\pm$0.60 & \nodata       &  \nodata & \nodata \\
              &      &  $H$ & 0.6 &  3.6 & $+$0.00$\pm$0.42  & $+$0.00$\pm$0.42  &  0.00$\pm$0.60 & \nodata       &  \nodata & \nodata \\
SA\,29$-$130  & DA   &  $J$ & 1.3 &  8.0 & $+$0.06$\pm$0.21  & $+$0.13$\pm$0.21  &  0.44$\pm$0.60 & 0.14$\pm$0.30  &  \nodata & \nodata \\
\enddata 
\tablenotetext{a}{Radius of aperture annulus for photometric measurements.}
\tablenotetext{b}{{Linear} polarization degree obtained with the ``difference'' method (telescope rotator angle of 0 deg).}
\tablenotetext{c}{{Linear} polarization degree obtained with the ``ratio'' method (telescope rotator angles of 0 and 90 deg).}
\tablenotetext{d}{Data from Whittet et al$.$ \cite{whittet92}.}
\tablenotetext{e}{This polarization angle is fixed at the literature value. We used it to find the polarization angle offset of our data (see text).}
\end{deluxetable}

\begin{deluxetable}{lcccccccccccc} 
\tabletypesize{\scriptsize}
\rotate
\tablecolumns{13} 
\tablewidth{0pc} 
\tablecaption{Linear polarization measurements of science targets\label{pol2}} 
\tablehead{ 
\colhead{Object}     & \colhead{SpT} & \colhead{Filt.}  & \colhead{FWHM}      & \colhead{Ann.\tablenotemark{a}} & \colhead{$q$}  & \colhead{$u$}  & \colhead{$P$\tablenotemark{b}} & \colhead{$P$\tablenotemark{c}} & \colhead{$\Theta$} & \colhead{$v$\,sin\,$i$} & \colhead{$P_{\rm rot}$} & \colhead{$E(J-[4.5])$}  \\
\colhead{(abridged)} & \colhead{}    & \colhead{}       & \colhead{(\arcsec)} & \colhead{(\arcsec)}             & \colhead{(\%)} & \colhead{(\%)} & \colhead{(\%)}                 & \colhead{(\%)}                 & \colhead{(deg)}    & \colhead{(km\,s$^{-1}$)} & \colhead{(h)}         & \colhead{(mag)}         \\
\colhead{(1)}        & \colhead{(2)} & \colhead{(3)}    & \colhead{(4)}       & \colhead{(5)}                   & \colhead{(6)}  & \colhead{(7)}  & \colhead{(8)}                  & \colhead{(9)}                  & \colhead{(10)}     & \colhead{(11)}          & \colhead{(12)}        & \colhead{(13)} 
}
\startdata 
J2244$+$20    & L7.5 & $H$ & 0.7                  & 4.2                  & $+$0.44$\pm$0.21                   & $+$0.43$\pm$0.21                   &  0.51$\pm$0.60                  & 0.62$\pm$0.30 &  \nodata   & $\le$30\tablenotemark{d} & 4.6                     & 1.33$\pm$0.26 \\
G\,196$-$3\,B & L3   & $J$ & 0.7\tablenotemark{e} & 4.2\tablenotemark{e} & $-$0.67$\pm$0.42\tablenotemark{e}  & $+$0.25$\pm$0.42\tablenotemark{e}  &  0.72$\pm$0.60\tablenotemark{e} & \nodata       &  \nodata   & 15.0                     & $\le$8\tablenotemark{g} & 1.12$\pm$0.23 \\
              &      & $J$ & 1.5\tablenotemark{f} & 2.3\tablenotemark{f} & $+$0.09$\pm$0.42\tablenotemark{f}  & $-$0.41$\pm$0.42\tablenotemark{f}  &  1.21$\pm$0.60\tablenotemark{f} & 0.42$\pm$0.60\tablenotemark{f} &  \nodata   &         &                         &               \\
USco\,128     & M7   & $J$ & 1.0 &  6.0 & $-$0.30$\pm$0.42  & $-$0.11$\pm$0.42  &  0.32$\pm$0.60 & \nodata       &  \nodata   & $\le$\,5                 & $\le$24\tablenotemark{g} & 0.78$\pm$0.22\tablenotemark{h} \\ 
              &      & $H$ & 1.1 &  6.0 & $-$0.17$\pm$0.42  & $+$0.09$\pm$0.42  &  0.19$\pm$0.60 & \nodata       &  \nodata   &                          &                          &                                \\
USco\,132     & M7   & $J$ & 1.0 &  5.0 & $-$0.21$\pm$0.42  & $-$0.22$\pm$0.42  &  0.31$\pm$0.60 & \nodata       &  \nodata   & \nodata                  & \nodata                  & 0.33$\pm$0.22\tablenotemark{h}  \\
J0136$+$09    & T2.5 & $J$ & 3.0 & 15.5 & $+$0.04$\pm$0.21  & $-$0.32$\pm$0.21  &  1.21$\pm$0.60 & 0.33$\pm$0.30 &  \nodata   & $\le$50\tablenotemark{d} & 2.4                      & 0.32$\pm$0.22 \\
J0241$-$03    & L0   & $J$ & 2.2 & 11.0 & $-$2.03$\pm$0.21  & $-$2.26$\pm$0.21  &  3.67$\pm$0.60 & 3.04$\pm$0.30 & 110$\pm$10 & \nodata                  & \nodata                  & 0.73$\pm$0.23 \\
J0355$+$11    & L5   & $J$ & 3.2 & 10.5 & $+$0.60$\pm$0.42  & $+$0.00$\pm$0.42  &  0.60$\pm$0.60 & \nodata       &  \nodata   & 12.4                     & $\le$10\tablenotemark{g} & 1.50$\pm$0.22 \\
J1022$+$58    & L1   & $J$ & 1.5 &  9.1 & $-$0.25$\pm$0.21  & $+$0.20$\pm$0.21  &  0.36$\pm$0.60 & 0.32$\pm$0.30 &  \nodata   & 15.0                     & $\le$8\tablenotemark{g}  & 0.03$\pm$0.22 \\
\enddata 
\tablenotetext{a}{Radius of aperture annulus for photometric measurements.}
\tablenotetext{b}{{Linear} polarization degree obtained with the direct measurement method (telescope rotator angle of 0 deg).}
\tablenotetext{c}{{Linear} polarization degree obtained with the flux ratio method (telescope rotator angles of 0 and 90 deg).}
\tablenotetext{d}{Derived from measured rotation periods (Section \ref{targets}). The true projected rotational velocity depends on the source's rotation axis inclination.}
\tablenotetext{e}{Data corresponding to 2006 Mar 23.}
\tablenotetext{f}{Data corresponding to 2011 Apr 20.}
\tablenotetext{g}{Rotation periods computed from $v$\,sin\,$i$ (Section \ref{targets}) and assuming $R = 0.1$ R$_\odot$.}
\tablenotetext{h}{The color $J-[4.5]$ is computed using the {\sl WISE} [4.6] magnitude (see text).}
\end{deluxetable}

\end{document}